\documentclass[%
 aip,
jap,%
bmf,%
 sd,%
rsi,%
 amsmath,amssymb,
preprint,%
reprint,%
]{revtex4-1}

\newcommand {\apgt} {\ {\raise-.5ex\hbox{$\buildrel>\over\sim$}}\ }
\newcommand {\aplt} {\ {\raise-.5ex\hbox{$\buildrel<\over\sim$}}\ }

\usepackage{graphicx}
\usepackage{dcolumn}
\usepackage{bm}

\begin{document}

\title{{Prediction of entropy stabilized incommensurate phases 
in the system $MoS_{2}-MoTe_{2}$.}}

\author{B. P. Burton }
\email{benjamin.burton@nist.gov}
\affiliation{Materials Measurement Laboratory,
National Institute of Standards and Technology (NIST),
Gaithersburg, MD 20899, USA}

\author{A. K. Singh}
\affiliation{Materials Measurement Laboratory,
National Institute of Standards and Technology (NIST),
Gaithersburg, MD 20899, USA}

\date{\today}

\begin{abstract}
A first principles phase diagram calculation, that included
van der Waals interactions, was performed for the 3D bulk
system $(1-X) \cdot MoS_{2} - (X) \cdot MoTe_{2}$.  
Surprisingly, the predicted phase diagram has 
at least two ordered phases, at $X \approx 0.46$, 
even though all calculated formation energies are positive; in a ground-state 
analysis that examined all configurations with 16 or fewer anion sites.
The lower-temperature {\bf $I$}-phase is predicted to transform to a 
higher-temperature {\bf $I^{\prime}$}-phase at $T \approx 500 K$, 
and {\bf $I^{\prime}$} disorders at $T \approx 730 K$. Both these 
transitions are predicted to be first-order, and there are broad two-phase
fields on both sides of the ordered regions. Both the {\bf $I$}- and
{\bf $I^{\prime}$}-phases are predicted to be $incommensurate$~ i.e. $aperiodic$: 
{\bf $I$}-phase in three dimensions; and {\bf $I^{\prime}$}-phase in two dimensions.

\end{abstract}


\maketitle

\section{Introduction}

Recently there has been great interest in two-dimensional (2D)
transition metal dichalcogenide (TMD) materials
such as $MoS_{2}$, $MoSe_{2}$~ and $MoTe_{2}$, their solid solutions, and
related 2D materials \cite{Wang,Ganatra2014}.  
Traditionally, $MoS_{2}$~ has been used as a dry lubricant \cite{DryLube} that 
is stable up to 623 K. Currently, interest is focused on applications as:
band-gap engineering materials \cite{Kang,Kutana}; 
nano-electronic devices \cite{Ganatra2014,Radisavljevic2011,Das2013,Wang2012}; 
photovoltaic devices \cite{Jariwala2014,Fontana2013}; 
valleytronics applications \cite{Zeng2012,Mak2012};
2D building blocks for electronic heterostructures \cite{2D}; and
as sensor materials \cite{Sensor}.  

The individual, three-atom-thick, 2D-layers of the bulk system are bonded 
by van der Waals forces, hence these forces influence bulk and multilayer 
synthesis and therefore anion order-disorder and/or phase separation 
in solid solutions.  The results presented below, for 3D bulk $MoS_{2}-MoTe_{2}$,  
imply that van der Waals interactions may strongly affect phase stabilities, 
either between adjacent layers in bulk or few-layer samples, or between 
monolayers and heterogeneous substrates.

Of the three quasibinary solid solutions ($MoS_{2}-MoSe_{2}$, 
$MoSe_{2}-MoTe_{2}$, $MoS_{2}-MoTe_{2}$) 
$(1-X) \cdot MoS _{2} -(X) \cdot MoTe _{2}$~
has the greatest difference in anionic radii 
($R_{S}$=1.84 \AA; $R_{Te}$=2.21 \AA) \cite{Radii}, which suggests
that it is the most likely to exhibit interesting solution behavior.
One expects a simple miscibility gap as 
reported by Kang et al. \cite{Kang} for
monolayer $MoS_{2}-MoTe_{2}$, hence the prediction of 
two configurational entropy ($S_{con}$) stabilized $incommensurate$, 
i.e. $aperiodic$, phases is extraordinary (stable phases that 
have positive formation energies must be entropy stabilized).

\section{Methodology}

\subsection{Total Energy Calculations}

Total structure energies, $\Delta E_{Str}$~ were calculated 
for fully relaxed $MoS_{2}$, $MoTe_{2}$ (2H-structure,
space group $P6_{3}/mmc$, AB-stacking of three-atom-thick layers), and for 233 
Mo$_{m+n}$(S$_{m}$Te$_{n}$)$_{2}$~ supercells. 
The Vienna $ab~initio$~ simulation program 
(VASP, version 5.3.3 ~\cite{Kresse1993,Disclaimer}) 
was used for all density-functional theory (DFT) calculations, with
projector augmented waves (PAW) and a generalized 
gradient approximation (GGA) for exchange energies. 
Electronic degrees of freedom were optimized with a conjugate gradient
algorithm. Valence electron configurations were: 
Mo\_pv$~4p^{5}5s4d$; S\_$s^{2}p^{4}$; Te\_$s^{2}p^{4}$. 
Van der Waals interactions that bond the three-atom thick
2D X-Mo-X layers (X=S, Te) together were modeled with the
non-local correlation functional of Klimes $et$~ $al.$ \cite{Klimes} 
Total energies were also calculated $without$~ van der Waals interactions,
but, up to a basis of 140 structures, ground-state analyses always
predicted false ground-states (supplementary material).
Convergence with respect to k-point meshes was achieved by increasing 
the number of k-points until the total energy converged.
A 500 eV cutoff-energy was used in the "high precision" option, 
which converges {\it absolute}~ energies to 
within a few meV/mol (a few tenths of a kJ/mol of exchangeable 
S- and Te-anions). Precision is at least an order of
magnitude better.
Residual forces of order 0.02 eV or less were typical.
Often, convergence with respect to hexagonal c-axis length was
not automatic, and it was necessary to chose an initial c-axis
value that is close to the converged value. Calculated interlayer spacings 
in $MoS_2$~ and $MoTe_2$ are 2.992 \AA~ and 3.513 \AA, 
respectively, corresponding experimental 
values are: 2.977 \AA~ \cite{Takeuchi} and 3.382 \AA~ \cite{Villars}.
     
\begin{figure}[!htbp]
\begin{center}
\vspace{-0.25in}
\includegraphics[width=7.0cm,angle=0]{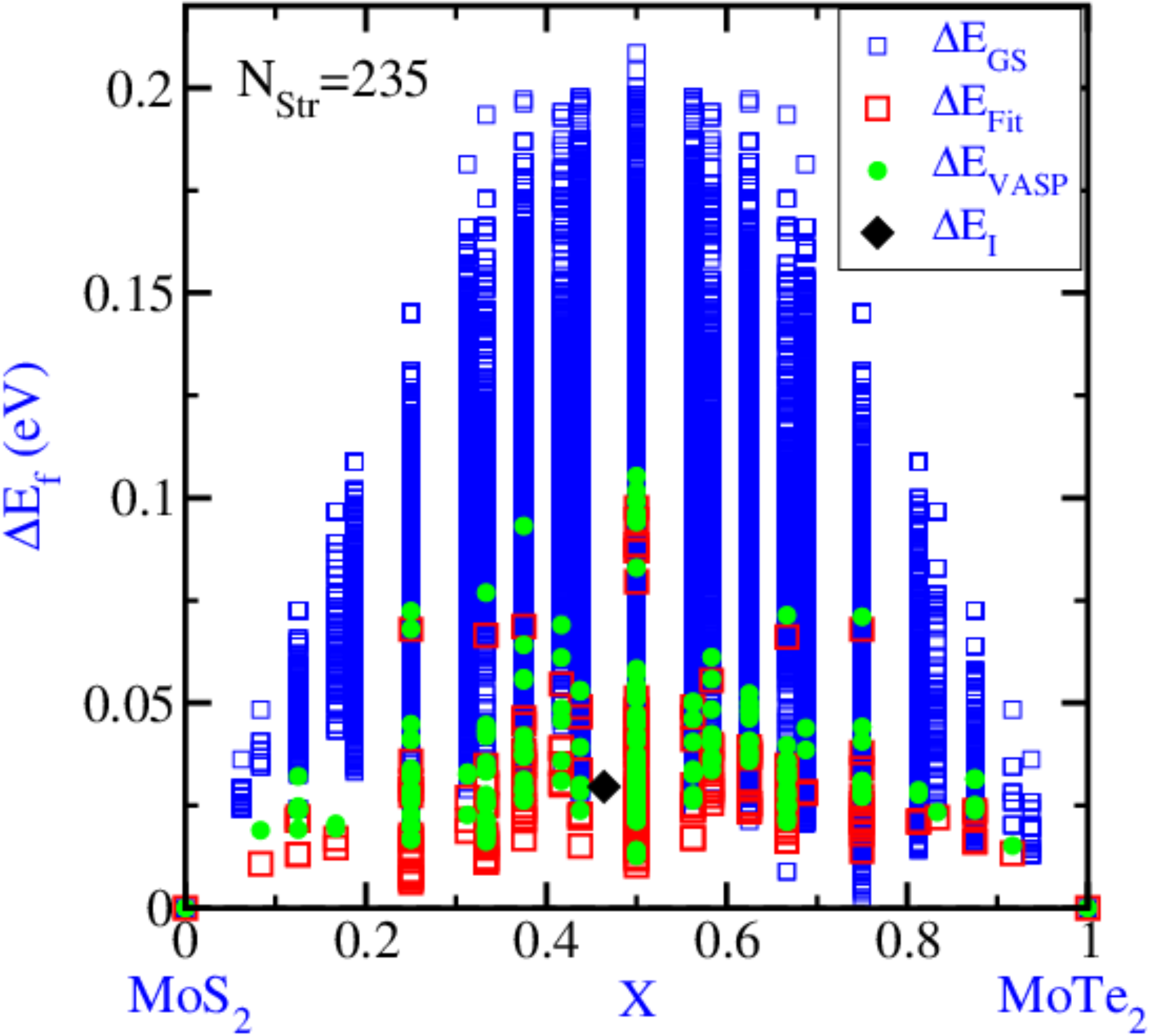}
\end{center}
\vspace{-0.20in}
\caption{ Comparison of formation energies,$\Delta E_{f}$, for the
235 DFT calculations (solid circles, green online) to Cluster Expansion (CE) formation energies: 
$\Delta E_{Fit}$~  (large open squares, red online) is the CE-fit to the DFT set; 
$\Delta E_{GS}$~ (smaller open squares, blue online) are the CE-based ground-state analysis;
$\Delta E_{I}$~ (solid black diamond at $X = 0.46$, $\Delta E_{I} \aplt 0.03$ eV) 
is the {\bf $I$}-phase formation energy.
All $\Delta E_{f}$\textgreater$0$~ implies that there are no ordered
ground-states, and suggests that the phase diagram will
have a miscibility gap. 
} 
\label{fg:eofx}
\end{figure}

Formation energies ($\Delta E_{f}$) for 233 Mo$_{l}$(S$_{m}$Te$_{n}$)$_{2}$~
supercells are plotted in Fig. \ref{fg:eofx}, in which 
values for $\Delta E_{f}$~ are normalized per mol of exchangeable  
anions, S and Te: 

\begin{equation}
\Delta E_{f} = (E_{Str} - mE_{MoS_{2}} - nE_{MoTe_{2}})/(2(m+n))
\end{equation}

\noindent
Here: $E_{Str}$~ is the total energy of the 
Mo$_{l}$(S$_{m}$Te$_{n}$)$_{2}$~ supercell; 
$E_{MoS_{2}}$  is the energy/mol of MoS$_{2}$; 
$E_{MoTe_{2}}$ is the energy/mol of MoTe$_{2}$.

All supercell energies are positive which suggests a 
miscibility gap system, unless one or more entropy 
stabilized phases are stable.

\subsection{The Cluster Expansion Hamiltonian}

A cluster expansion Hamiltonian (CEH) \cite{Sanchez1984}, 
for the (1-X)$\cdot$MoS$_{2}$-(X)$\cdot$MoTe$_{2}$ quasibinary system was 
fit to the set of 235 formation energies, $\Delta E_{VASP}$, solid dots
(green online) in Fig. \ref{fg:eofx} with a cross validation score of
(CV )2=0.00723896).
Fitting of the CEH was performed with the Alloy Theoretic Automated Toolkit (ATAT)
\cite{Disclaimer,Axel2002a,Axel2002b,Axel2002c} which automates most of the tasks
associated with CEH construction. A complete description
of the algorithms underlying the code can be found in
\cite{Axel2002b}. 
Large open squares in Fig. \ref{fg:eofx} (red online) indicate values of 
the 235 $\Delta E_{Fit}$~ that were calculated with the CEH. Smaller open squares
($\Delta E_{GS}$, blue online) indicate the results of a ground-state analysis in which
the CE was used to calculate formation energies for all ordered configurations
with 16 or fewer anion sites, 151,023 structures.  

\section{Results and Discussion}
\begin{figure}[!htbp]
\begin{center}
\vspace{-0.2in}
\includegraphics[width=6.9cm,angle=0]{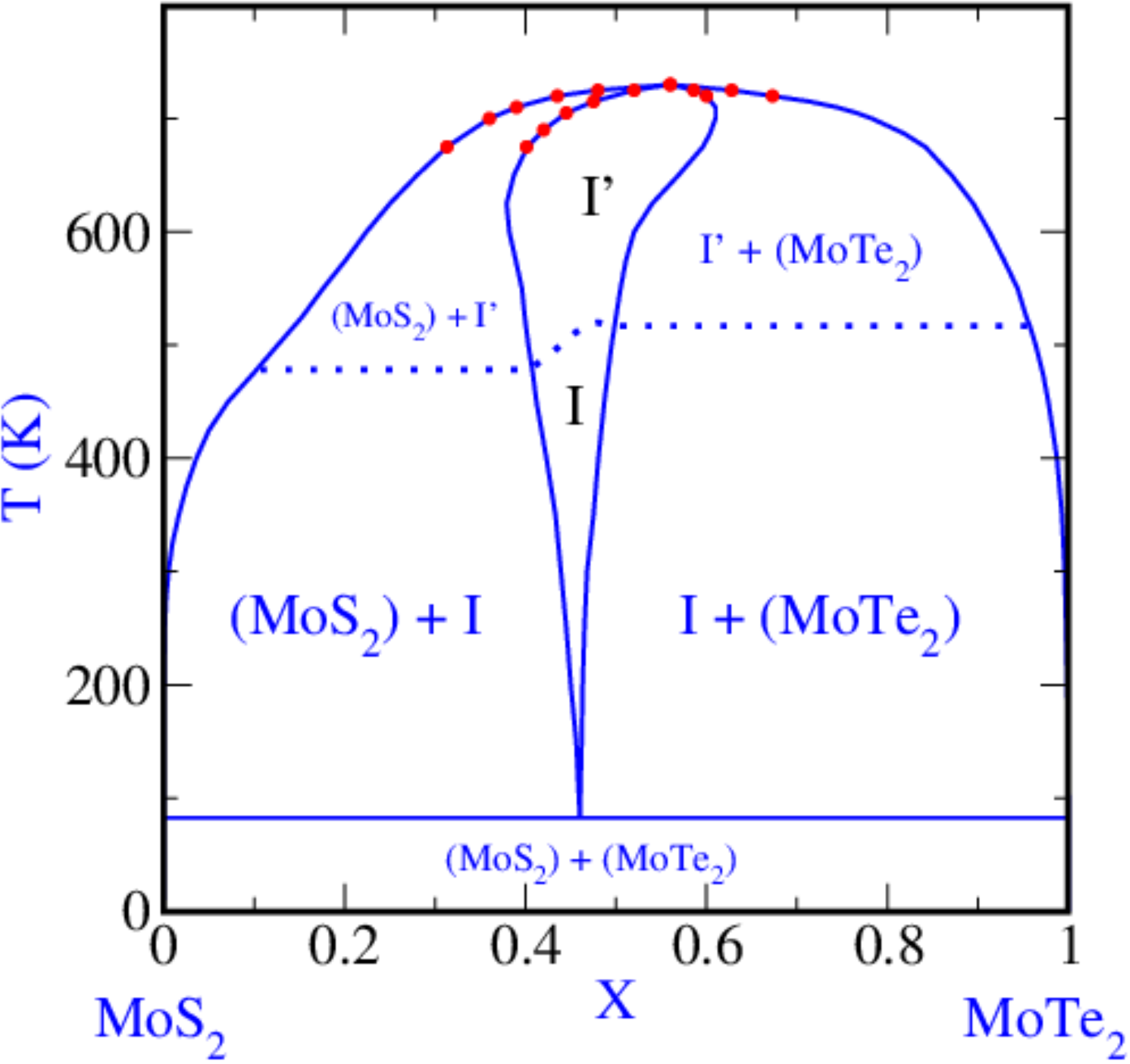}
\end{center}
\vspace{-0.2in}
\caption{ Calculated phase diagram. The isothermal
line at 82.5 K indicates that the {\bf $I$}-phase, 
at $ X \approx 0.46$, is entropy stabilized (not a ground-state); 
Here: $(MoS_{2})$~ and $(MoTe_{2})$ indicate $MoS_{2}$-rich
and $MoTe_{2}$-rich solid solutions, respectively. Dotted lines
in the region of the {\bf $I^{\prime }  \rightleftharpoons  disordered$} ~
transition indicate inferred extensions of calculated phase boundaries.
}
\label{fg:XT}
\end{figure}

A first principles phase diagram (FPPD) calculation was performed
with grand-canonical, and canonical, Monte Carlo (MC) simulations using the 
emc2 and phb codes which are part of the ATAT
package \cite{Axel2002a,Axel2002b,Axel2002c}. 
Most phase boundaries were calculated with the phb program
which uses equilibration tests to set the numbers of 
equilibration- and MC-passes \cite{Axel2002c}. 
To draw high-T extensions of the two-phase
fields, and to locate the $I \rightleftharpoons  I^{\prime }$~ transition, 
a 48x48x12 unit cell simulation box box was used, 
with 2000 equilibration passes and 2000 MC-passes (see suplimentary
material for comparisons of various equilibration- and MC-pass settings
in calculations of the {\bf $I \rightleftharpoons  I^{\prime }$} 
phase transition). 
The predicted phase diagram is shown in Fig. \ref{fg:XT};
where $(MoS_{2})$~ denotes an $MoS_{2}$-rich solution phase,
and similarly for $(MoTe_{2})$.

\vspace{+0.2in}

Kang et al.\cite{Kang} performed first principles phase diagram
calculations for four dichalcogenide monolayer systems:
$MoSe_{2(1-x)}Te_{2x}$, $WSe_{2(1-x)}Te_{2x}$,
$MoS_{2(1-x)}Te_{2x}$ and $WS_{2(1-x)}Te_{2x}$.
van der Waals interactions, and fitting 
Their CEs wer fit to about 40 structures per system, and 
van der Waals interactions (with substrate) omitted. 
All systems were predicted to have miscibility gaps, and surprisingly,
all consolute points are on the Te-rich sides.
One expects the consolute point to be on
the S-rich side, because it typically requires less
energy to substitute a smaller S-ion into
a larger Te-ion site, than vice versa.
Figure \ref{fg:XT} also has reduced solubility on the Te-rich side,
but this is related to immiscibility between the {\bf I}-
({\bf $I^{\prime}$})-phase, and the Te-rich phase.

It is not clear that the Kang et al. $MoS_{2}-MoTe_{2}$~ phase diagram 
would still be a simple miscibility gap had they included hundreds of 
structures in their CE-fit rather than about 40. Hence the monolayer vs. bulk 
comparison is uncertain.

Surprisingly, the phase diagram predicted here has multiple 
two-phase fields, separated by two ordered incommensurate phases, 
neither of which is a ground state. 
To seven digits, the calculated bulk composition of the {\bf I}-phase, 
just above its 82.5 K minimum temperature of stability, is X=0.4642857=13/28; 
i.e. $Mo_{14}S_{15}Te_{13}$. 
Stability of the {\bf I}-phases is a robust result, in Monte-Carlo
simulations: (1) CEH fits to 128, 153, 162, 182, 225 and 235 formation energies all
predict {\bf I}-type ordering; (2) {\bf I}-phase forms spontaneously on cooling
of the {\bf $I^{\prime }$}-phase at $T \aplt 500 K$, and on heating 
of a low-T equilibrium $(MoS_{2})~+~(MoTe_{2})$~ assemblage to $T \apgt 82 K$;
(3) {\bf I$^{\prime }$}-phase forms spontaneously on 
heating of the {\bf I}-phase, or cooling of a disordered solid solution 
with $X \approx 0.46$.  This calculation considers only $S_{con}$, and ignores 
excess vibrational entropy, $S_{vib}$, which could conceivably
destabilize the {\bf I}-phases. In light of (1) above, however, this seems 
highly unlikely.  Also, there is no fully satisfactory way to model $S_{vib}$~ 
for an aperiodic phase, and a reasonable approximate structure
(with {\bf I}-phase like ordering) would require at least a low symmetry 
84-atom cell; which is beyond the scope of this study.     

\vspace{+0.2in}
\begin{figure}[!htbp]
\begin{center}
\vspace{-0.20in}
\includegraphics[width=6.0cm,angle=0]{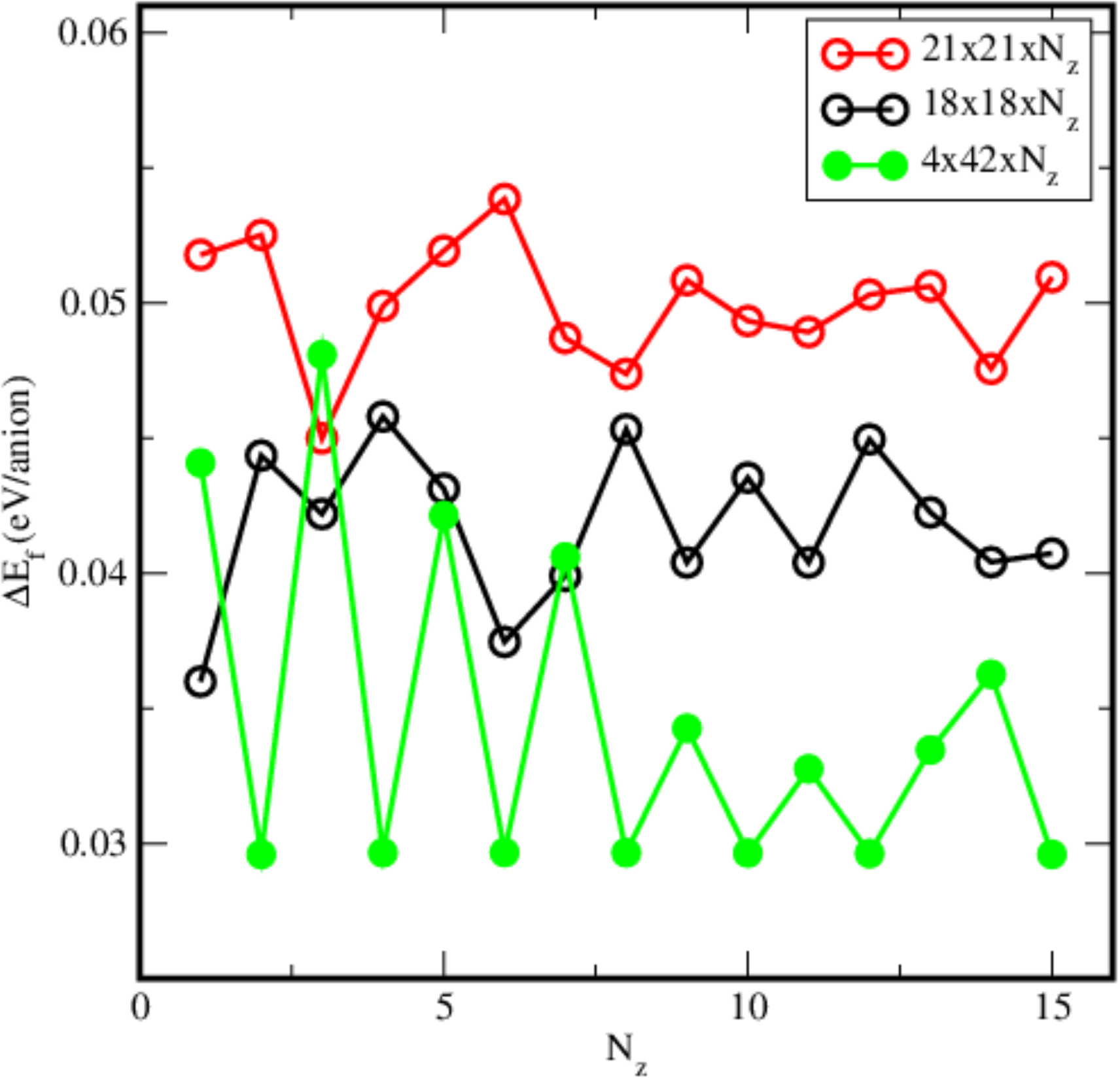}
\end{center}
\vspace{-0.3in}
\caption{Minimum energies for various Monte-Carlo supercells; N$_z$~
is the length of the supercell, in c-axis units of the $P6_{3}/mmc$~ 
disordered-phase cell constants.  The flat minimum at 
$\Delta E_{f} \equiv \Delta E_{I} \aplt 0.03$ eV/anion is interpreted as
the {\bf $I$}-phase formation energy. 
}
\label{fg:NotGS}
\end{figure}

\begin{figure}[!htbp]
\begin{center}
\includegraphics[width=7.0cm,angle=0]{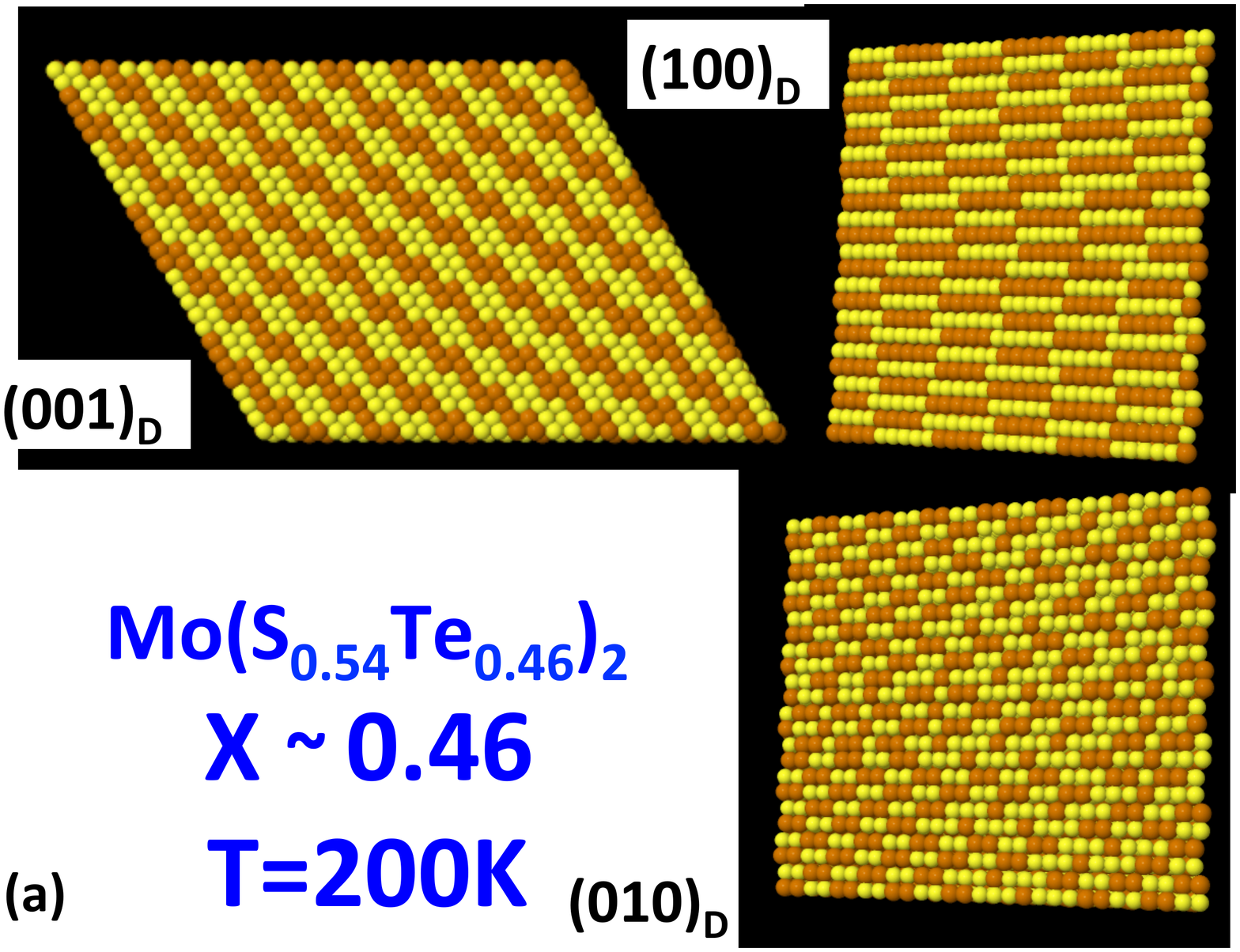} 
\includegraphics[width=7.0cm,angle=0]{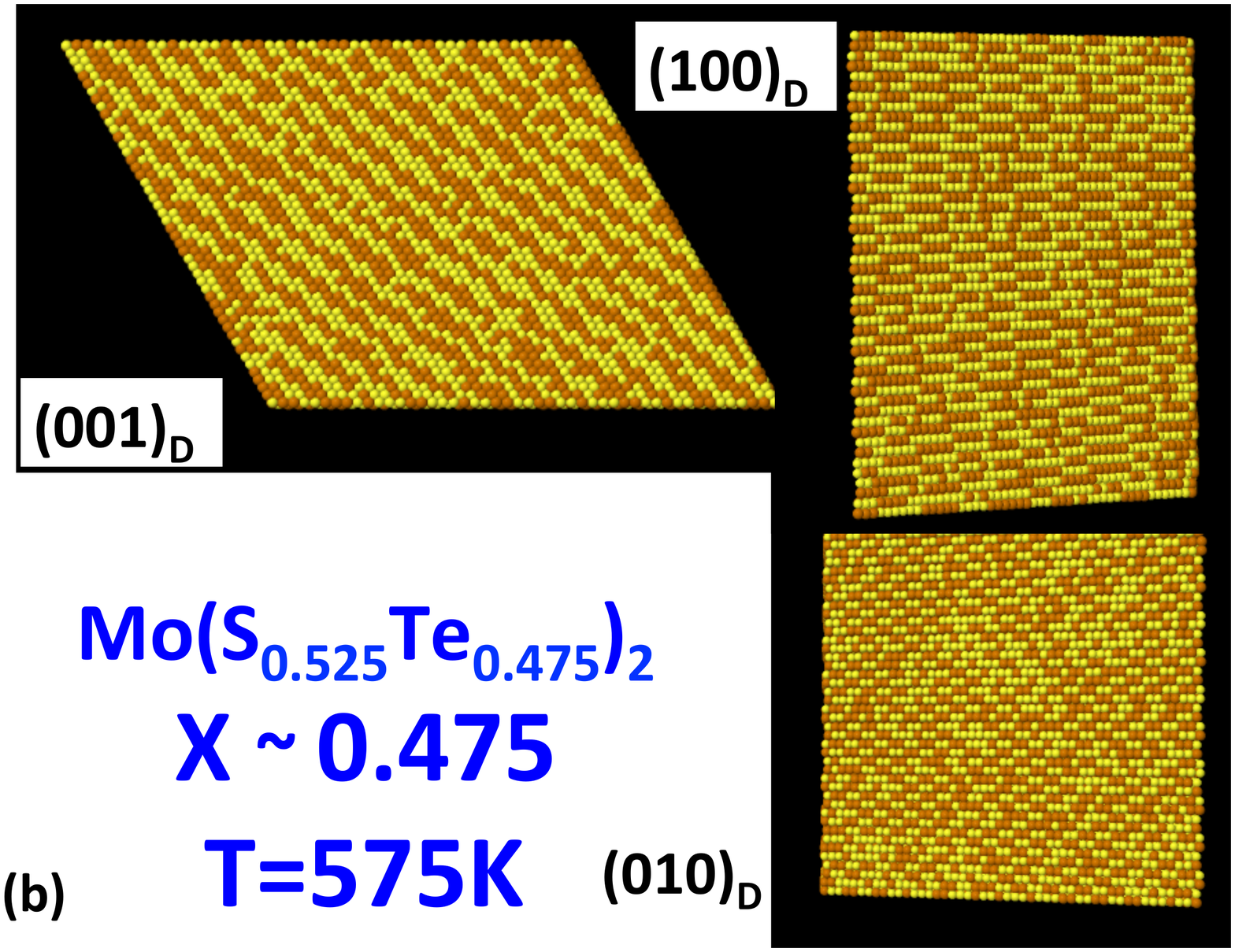}
\end{center}
\vspace{-0.2in}
\caption{ Monte-Carlo-snapshots of S:Te-ordering in the {\bf $I$}- and 
{\bf $I^{\prime}$}-phases at: 
(a) $ X \approx 0.46$~ and T=200 K; and 
(b) $ X \approx 0.475$~ and T=575 K, respectively   
(online S=yellow, Te=brown, Mo omitted for clarity). 
Labels (001)$_D$, (100)$_D$~ and (010)$_D$, refer to 
corresponding crystallographic planes in the high-T 
$P6_{3}/mmc$~ disordered phase. 
}
\label{fg:ORD}
\end{figure}

Figure \ref{fg:NotGS} shows how the CE-calculated
{\bf $I$}-phase formation energy $\Delta E_{f}(I-phase) \equiv \Delta E_{I}$, 
varies as a function of MC-supercell size and shape. The flat minimum 
at $\Delta E_{I} \aplt 0.03$ eV/anion is the calculated 
{\bf $I$}-phase formation energy which is plotted as 
the solid black diamond in Fig. \ref{fg:eofx}.  
Supercell dimensions were chosen to accommodate $Mo_{14}S_{15}Te_{13}$~
stoichiometry and {\bf I}-phase ordering.  Note that many of 
the $\Delta E_{f}$~ plotted in Figs. \ref{fg:eofx}, are lower 
in energy than $\Delta E_{I}$, but that they are 
for $periodic$~  structures with 16 or fewer anion sites in 
which $S_{con} \rightarrow 0$~ as $T \rightarrow 0 K$, and 
clearly (Figs. \ref{fg:ORD}) the {\bf $I$}-phase is incommensurate, 
i.e. $aperiodic$~ 
with $S_{con}$ $>$ 0.  Figures \ref{fg:ORD} 
exhibit the S:Te (yellow:brown online, respectively) ordering 
at: (a) 200 K; and (b) 575 K, i.e. below and above 
the {\bf $I \rightleftharpoons  I^{\prime }$} phase transition. 
Mo-atoms are omitted for clarity, and 
labels (001)$_D$, (100)$_D$, and (010)$_D$~ refer to corresponding 
crystallographic planes in the high-T $P6_{3}/mmc$~ $disordered$-phase. 

In the (001)$_D$- and (100)$_D$-planes, ...$S_{m}Te_{n}$... chains in the  
$<$010$>_D$~ direction, most often have $m=5~or~6$~ and $n=4~or~5$.
Also, in (100)$_D$, the ...$S_{m}Te_{n}$... chains exhibit irregular alignments
relative to one another.  Note however, that $S_{m}Te_{n}$-chains in 
(001)$_D$-planes {\bf $order$} along the $<$001$>_D$~ direction, 
such that $S_{m}$-units alternate with
$Te_{n}$-units in adjacent 3-atom thick 2D-layers; i.e.  
...$S_{m}Te_{n}S_{m^{\prime }}Te_{n^{\prime }}$...chains are stacked on top of 
...$Te_{n}S_{m}Te_{n^{\prime }}S_{m^{\prime }}$...chains with
inescapable misfits, owing to the different and variable values 
of $m$~ and $n$. Thus {\bf $I$}-phase ordering is inevitably 
imperfect, aperiodic, and incommensurate, which suggests that
the {\bf $ I \rightleftharpoons  I^{\prime } $} phase transition
is first-order.

The difference between {\bf $I$}- and 
{\bf $I^{\prime }$}-phases appears to be a distinction between 
3D-ordering in the low-T {\bf $I$}-phase
and 2D-ordering in the high-T {\bf $I^{\prime }$}-phase.
Clearly, ordering is stronger in the $basal$~ (001)$_D$-plane than in
the (100)$_D$- or (010)$_D$-planes, and striped order within
(001)$_D$~ persists above the 
{\bf $I \rightleftharpoons  I^{\prime }$} transition. 

\begin{figure*}
\begin{center}
\vspace{-0.20in}
\includegraphics[width=16.0cm,angle=0]{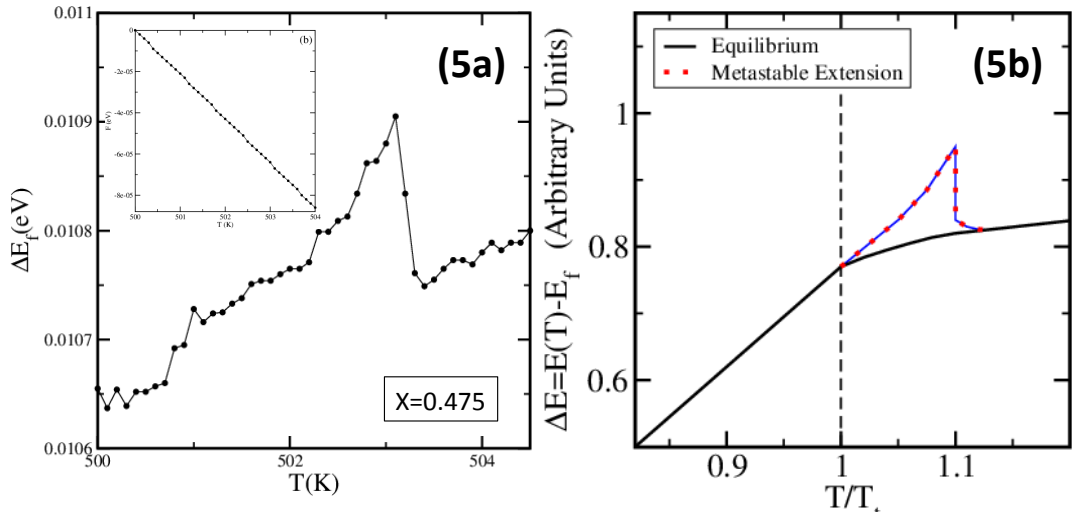}
\end{center}
\vspace{-2.0in}
\caption{ a) Monte-Carlo T-scans (heating) of: total energy, $E_{TOT}(T)$, 
and Helmholtz energy, $F(T)$, (inset) as functions of temperature;
b) idealized schematic comparing equilibrium- and metastable transition
paths (solid black line and red-dotted blue line, online, respectively).  
$E_{TOT}(T)$~ indicates a phase transition at $T \approx 500 K$; 
the absence of a clear change of slope in $F(T)$~ indicates 
that the transition is only weakly
first-order; hence the dotted {\bf $I \rightleftharpoons  I^{\prime }$}~
transition line in Fig. \ref{fg:XT}.
}
\label{fg:trans}
\end{figure*}

Figure \ref{fg:trans}a is a Monte-Carlo T-scan (heating) of the 
total energy $E_{TOT}(T)$~ which confirms first-order character for
the {\bf $I \rightleftharpoons  I^{\prime }$} transition: a critical
(continuous) transition \cite{Fisher} would not exhibit 
the sharp change at $T\approx503 K$; 
also a transition from the lower-T, higher-$E_{TOT}$~ phase to the 
higher-T, lower-$E_{TOT}$~ phase requires that the the lower-T 
phase be superheated, i.e. $metastable$~ before it transforms.
Figure \ref{fg:trans}b is an idealized schematic that compares
equilibrium- and metastable-transition paths (solid black line,
and blue line with red dots, online respectively). This transition
is more subtle in cooling simulations, but still evident in snapshots.
In the Fig. \ref{fg:trans} inset, no change of slope at the transition 
is evident in the Helmholtz energy, F(T), which suggests that the
transition is $weakly$~ first-order \cite{note}. Also, when the
MC temperature-increment was decreased from 1.0 K/MC-step to 0.1 K/MC-step (not shown), 
the predicted transition temperature decreased by about 10 K, which 
indicates more superheating at 1.0 K/MC-step than at 0.1 K/MC-step, hence 
first-order character.  This transition is shown
as a dotted line in Fig. \ref{fg:XT} because the two-phase fields
that a first-order transition implies are too narrow to
resolve in the MC-simulations.


\section{Conclusions}

To summarize, a first principles phase diagram calculation for the 3D bulk system
$(1-X) \cdot MoS _{2} -(X) \cdot MoTe _{2}$, that includes van der Waals
interactions, predicts the formation of two entropy stabilized 
$incommensurate$, i.e. $aperiodic$~ phases: the {\bf $I$}- and 
{\bf $I^{\prime }$}-phases, at X$ \approx 0.46$. 
Above the minimum temperature for stability of the {\bf $I$}-phase, 
$T \approx 82 K$, the calculation predicts broad two-phase fields between 
the {\bf $I$}- or {\bf $I^{\prime }$}-phase and disordered 
S- or Te-rich solution phases, ($MoS_{2}$) and ($MoTe_{2}$), 
respectively.  Both the {\bf $I \rightleftharpoons  I^{\prime }$}~ and 
{\bf $I^{\prime }  \rightleftharpoons disordered$}~ 
transitions are predicted to be first-order. Dramatic changes in
phase relations can be induced by arbitrarily small differences in energy, 
hence van der Waals interactions should not be ignored in layered 2D-systems
such as $MoS_{2} - MoTe_{2}$.


\section*{ACKNOWLEDGEMENTS}
This work was supported by NIST-MGI, and B. Burton wishes to thank
R. Selinger, J. Douglas, K. Migler, A. Davydov, B. Campbell, J. Lau and L. Bendersky for
useful discussions.


\begin{thebibliography}{999}
\bibitem{Disclaimer} 
Reference to specific software packages does not imply
a NIST endorsement.
\end{thebibliography}
\end{document}



\section{SUPPLEMENTARY MATERIAL}

\begin{figure}[!htbp]
\begin{center}
\vspace{-0.10in}
\includegraphics[width=14.0cm,angle=0]{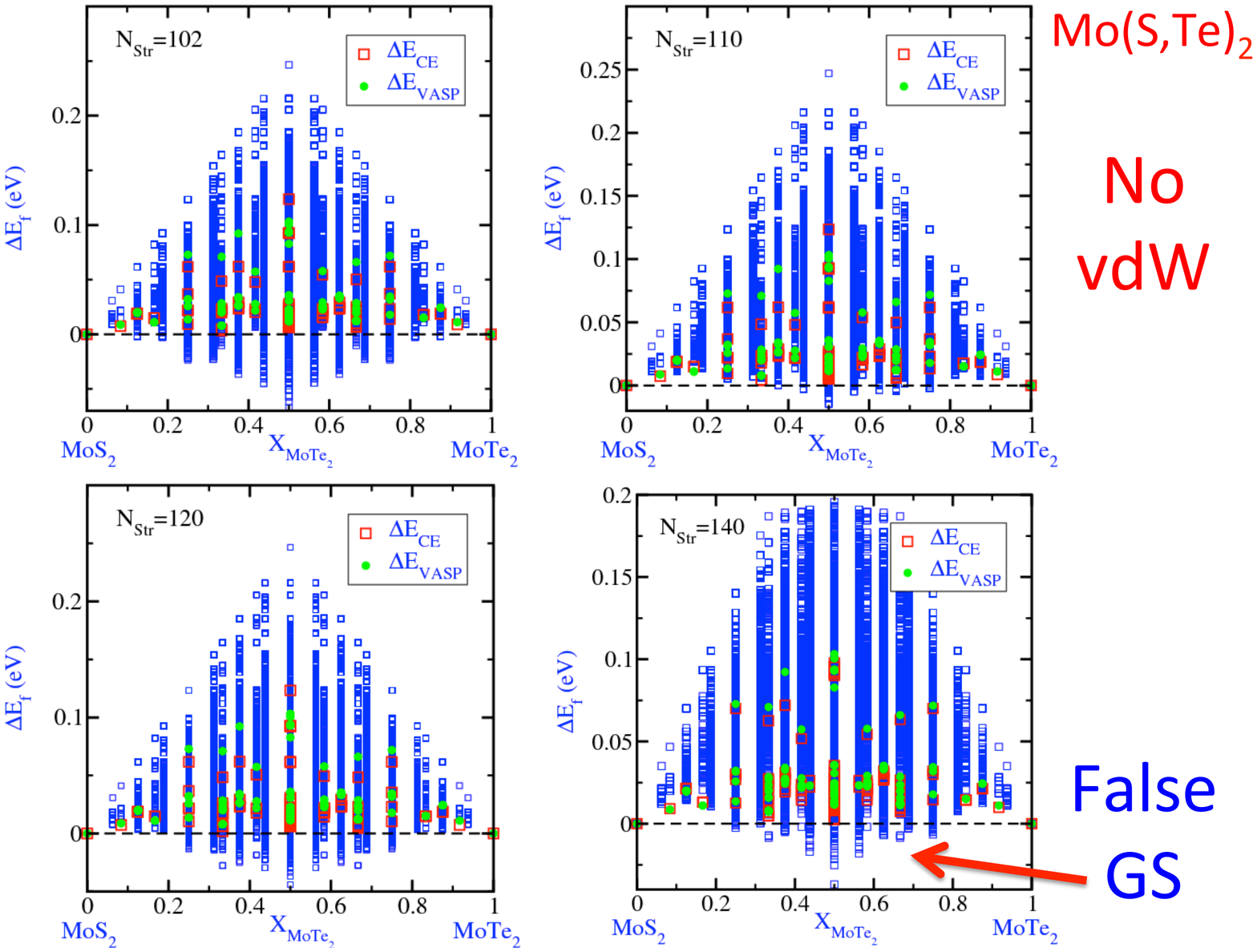}
\end{center}
\vspace{-0.25in}
\caption{Ground state analyses for cluster expansions based on
fits to 102, 110, 120 and 140 formation energies, that
were calculated without including van der Waals interactions.
Note the presence of false ground-states in each fit; i.e. blue 
squares below the zero-energy line.  
}
\label{fg:NOvdW}
\vspace{-0.20in}
\end{figure}

\begin{figure}[!htbp]
\begin{center}
\vspace{-0.10in}
\includegraphics[width=14.0cm,angle=0]{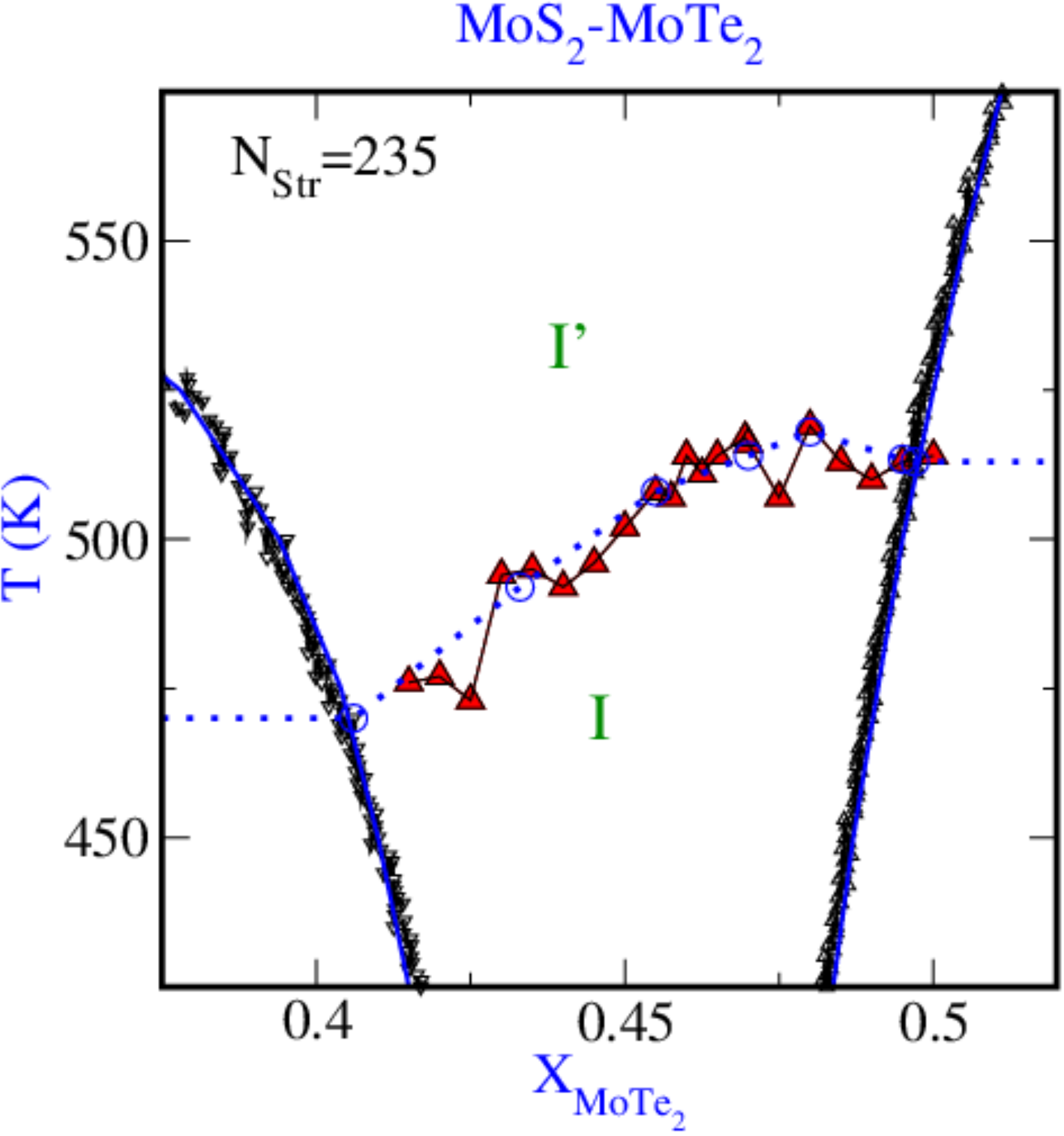}
\end{center}
\vspace{-0.25in}
\caption{Calculated $I \rightleftharpoons  I^{\prime }$~ transition
points (solid up-pointing triangles, red online). Scatter in these
results reflect MC-simulation fluctuations. The phase boundaries,
$(MoS_2)+I(I^{\prime }) \rightleftharpoons I(I^{\prime })$~ and
$I(I^{\prime }) \rightleftharpoons I(I^{\prime })+(MoTe_2)$, were
calculated with the phb program in the ATAT package. 
}
\label{fg:PDataII}
\vspace{-0.20in}
\end{figure}

\begin{figure}[!htbp]
\begin{center}
\vspace{-0.10in}
\includegraphics[width=14.0cm,angle=0]{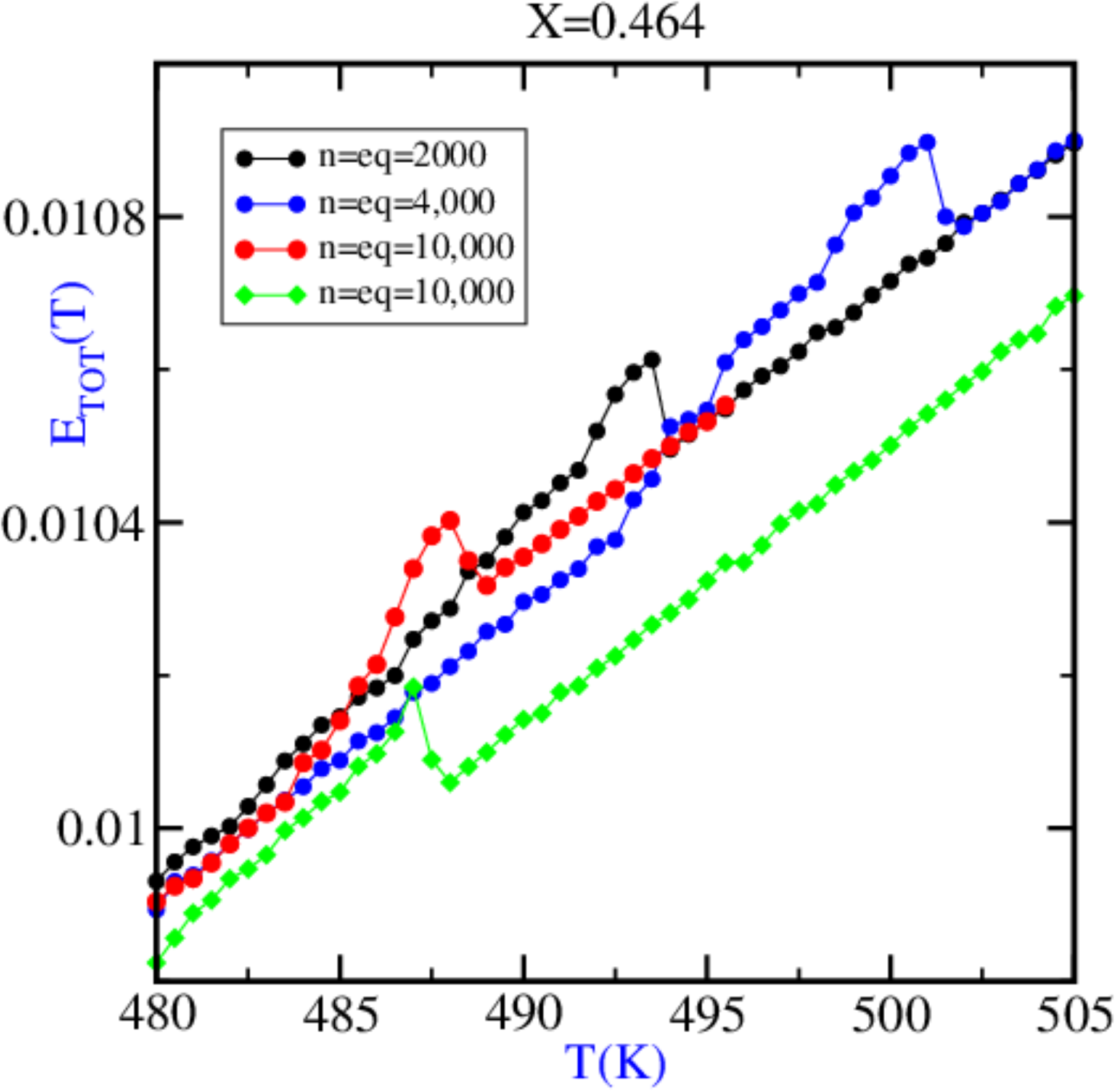}
\end{center}
\vspace{-0.25in}
\caption{Calculated $I \rightleftharpoons  I^{\prime }$~ transition
E$_{TOT}$(T) curves as functions of the numbers of MC- and 
equilibration-passes (n and eq, respectively) with a 48x48x12 
MC-simulation box (solid circles), and a 56x56x12 MC-box (solid diamonds).
}
\label{fg:E464}
\vspace{-0.20in}
\end{figure}